\documentclass[trackchanges,twocolumn]{aastex7}
\usepackage{multirow}
\usepackage{diagbox}
\usepackage{booktabs}

\begin{document}

\nolinenumbers
\title{Unified Formation Channel of Hot and Warm Jupiters via Planet-Planet Scattering}

\author[orcid=0009-0001-3775-7934, sname='Julia Esposito']{Julia Esposito}
\affiliation{Georgia Institute of Technology, School of Physics}
\email[show]{jesposito33@gatech.edu}  

\author[orcid=0000-0001-8308-0808, gname='Gongjie', 
sname='Li']{Gongjie Li} 
\affiliation{Georgia Institute of Technology, School of Physics}
\email{gongjie.li@physics.gatech.edu}

\author[orcid=0000-0002-7846-6981, gname='Songhu', 
sname='Wang']{Songhu Wang} 
\affiliation{Indiana University, Department of Astronomy}
\email{sw121@iu.edu}

\begin{abstract}

Recent observations show distinct orbital architectures for hot and warm Jupiters: hot Jupiters span a wide range of stellar obliquities and tend to host distant companions without close-by companions, whereas warm Jupiters are often aligned and accompanied by both close-by and distant companions. In this paper, we revisit planet–planet scattering and demonstrate that it provides a unified framework for both populations. Using $N$-body simulations with tides, we explore three regimes: hot ($a_1<0.1$ AU), warm ($0.1<a_1<1$ AU), and cold ($1<a_1<10$ AU) scattering. Hot scattering predominantly produces compact hot-Jupiter pairs, which are rarely observed, implying this channel is rare. Cold scattering readily produces retrograde hot Jupiters and likely constitutes a main reservoir feeding the hot-Jupiter population. However, cold scattering produces few inner warm Jupiters at $a\simeq0.1$–$0.3$ AU. We show that warm scattering naturally fills this gap: high-inclination inner warm Jupiters produced by warm scattering are preferentially removed through further eccentricity excitation followed by tidal circularization into hot Jupiters. As a result, the surviving inner warm Jupiters are biased toward a broad range of eccentricities but modest inclinations, producing the observed “eccentric-but-aligned” population. This story makes testable predictions: (i) warm Jupiters, especially at $a\gtrsim0.3$ AU, should not be exclusively aligned, and (ii) warm Jupiters should often host nearby companions with non-negligible mutual inclinations up to $\lesssim 30^\circ$.

\end{abstract}


\keywords{\uat{Extrasolar gaseous planets}{2172} --- \uat{Exoplanet astronomy}{486} --- \uat{Exoplanet formation}{492} --- \uat{Exoplanet migration}{2205} --- \uat{Hot Jupiters}{753}}

\section{Introduction}

Observationally, hot Jupiters exhibit a broad range of stellar obliquities \citep[as reviewed by][]{Winn15, Triaud2018, Albrecht22}, and are often dynamically isolated \citep{Steffen12, Huang16, Wu23}, whereas warm Jupiters are typically more aligned \citep{Rice22_1, Wang24}, and are more commonly found in compact multi-planet architectures \citep{Huang16, Wu23}. Taken at face value, these contrasts have often been interpreted as evidence of distinct formation pathways for the hot- and warm-Jupiter populations.

However, several subtler observational clues suggest a more complicated picture. Although hot Jupiters are dynamically hot in aggregate, the small fraction that retain nearby companions point to a comparatively quiescent dynamical history in at least some systems \citep[e.g.,][]{Becker15, Canas19, Wang21, Wu23}. Conversely, although warm Jupiters are primarily dynamically cool, a subset reach very large eccentricities that are difficult to reconcile with a purely gentle origin without additional dynamical excitation \citep{Goldreich03}.

No single standard hot-Jupiter origin model --- disk-driven migration \citep{Goldreich80, Lin86}, high-eccentricity migration \citep{Rasio96, Wu11, Chatterjee08, Naoz11, Petrovich15, Noaz16}, or \textit{in-situ }formation \citep{Batygin16} --- straightforwardly reproduces this mixture of dynamically ``hot-and-cool" properties across both populations. In this paper, we revisit planet–planet scattering as a unified dynamical framework capable of producing both hot and warm Jupiters while naturally yielding their distinct architectural outcomes.

Planet-planet scattering is a natural outcome of planets migrating in the protoplanetary disk\citep{Kokubo02, Goldreich04, Ida04}. Specifically, planets migrate close to each other in a compact configuration within the disk, and as the disk disperses, the planets perturb each other's orbits more significantly \textbf{\citep{Matsumura10}}, leading to instability and scattering. 

Historically, planet–planet scattering experiments have been conducted mainly and extensively in the cold-Jupiter region, and they have successfully explained the origin of misaligned hot Jupiters, as well as wide orbit planets \citep{Rasio96, Chambers96, Lin97, Adams03, Boss06, Ford08, Chatterjee08, Scharf09, Veras09, Mustill17}. This focus reflects a Solar System-centric expectation in which giant planets form primarily beyond the snow line. To leading order, this view is consistent with occurrence-rate constraints, since cold Jupiters are an order-of-magnitude more common than hot and warm Jupiters \citep{Fulton2021, Wittenmyer2020}. However, it does not exclude the possibility that multiple giant planets can exist at smaller radii, which would naturally motivate dynamical instabilities operating \textit{in situ} over a wider range of orbital distances, including the hot- and warm-Jupiter regions \citep{xu2024}.

When scattering occurs in the hot Jupiter region ($0.05~\mathrm{AU} < a < 0.15~\mathrm{AU}$), \citet{Petrovich14} found that collisions were the dominant outcome due to a low Safronov number, and the resulting planets are unlikely to achieve high eccentricities and inclinations.  In the warm Jupiter region, \citet{Juric08}, \citet{Frelikh19}, \citet{Anderson19}, it is found that scattering can lead to moderate eccentricities in agreement with observations. It remains unclear, however, whether the inclination (or spin–orbit) distribution produced by warm-scattering is consistent with observations, given that observed warm Jupiters --- including many eccentric systems --- tend to be spin–orbit aligned \citep{Wang24, Espinoza-Retamal2023, Rice22_1}.

In this paper, we re-examined planet-planet scattering across the hot, warm and cold regions ($0.03-10$AU) as a contributor to the formation of both warm and hot Jupiters. Additionally, we included a simple prescription for tides to circularize and shrink the planetary orbits. We tested whether planet-planet scattering can explain both the population of observed moderately eccentric, low-inclination warm Jupiters as well as the broad distribution of misalignments in hot Jupiters. 
This paper is organized as follows: we begin by describing the setup of our scattering experiment in section \ref{sec:setup}. Then, we discuss our results on hot Jupiters (section \ref{sec:hotjup}), warm Jupiters (section \ref{sec:warmjup}), as well as their companions (section \ref{sec:companions}). We conclude with a discussion and predictions for future observations in section \ref{sec:summary}.

\begin{figure*}[htbp]
    \centering    \includegraphics[width=0.7\textwidth]{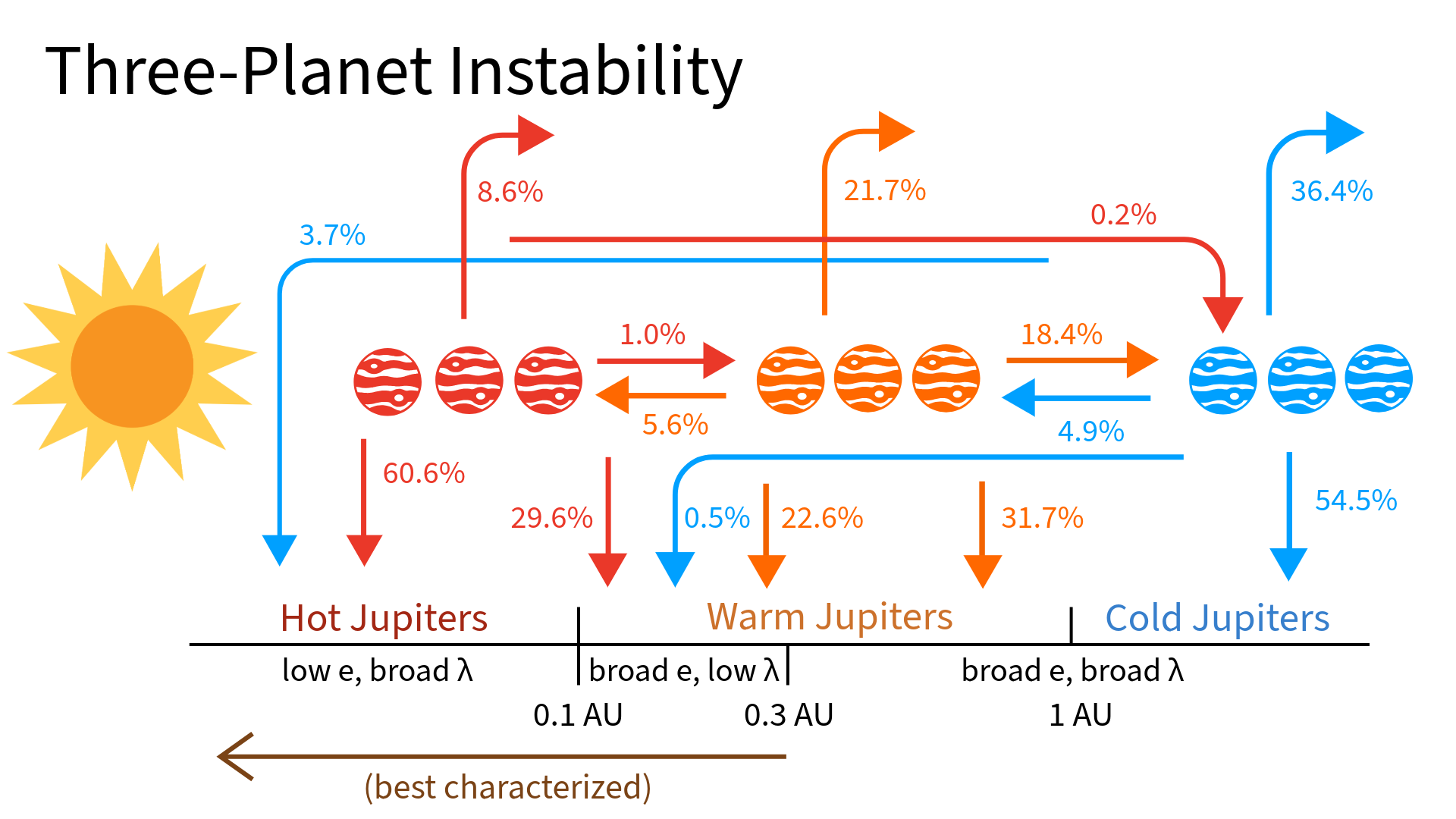}
    \caption{A visualization of pathways from hot, warm, and cold scattering. Each percentage is out of the total number of planets at the end of the simulation. Planets involved in collisions are only counted as one planet. The range of 0.1 to 0.3 AU shows the region that best characterizes and defines a warm Jupiter. We refer to these planets as "inner warm Jupiters." This range for warm Jupiters is used in Figure \ref{fig:a_i_e}.}
    \label{fig:cartoon}
\end{figure*}

\section{Setup of Scattering Experiments}
\label{sec:setup}
For the scattering simulations, we considered systems consisting of three planets orbiting a solar-mass host star. As shown in \citet{Chatterjee08}, the initial conditions prior to reaching 2-3 planets do not matter as the system has essentially ``forgotten'' the initial conditions due to the chaos. Therefore, it is reasonable for scattering experiments to start with two to three planets. The initial semi-major axis of the innermost planet was drawn log-uniformly between 0.03 and 0.1 AU for the hot scattering experiments, 0.1 and 1 AU for the warm scattering experiments, and between 1 and 10 AU for the cold scattering experiments. We split the warm Jupiter region into two sub-regions when we analyze the results: inner warm Jupiters ($0.1 < a < 0.3 $ AU), which are currently better characterized observationally, and outer warm Jupiters ($ 0.3 < a < 1 $ AU). 

The semi-major axes of the other two planets were then assigned based on their mutual Hill radius separations. Specifically, the semi-major axes of the second and third planets were determined by:
\begin{equation}
a_i = KR_{H,\mathrm{mut}} + a_{i-1}
\end{equation}
where
\begin{equation}
R_{H,\mathrm{mut}} = \frac{1}{2} (a_{i-1}+a_i) (\frac{m_{i-1}+m_i}{3M_*})^{1/3}
\end{equation}
and $K = 4$. 
In Figure \ref{fig:cartoon} we illustrate the results of the planet-planet scattering set-up starting from three planets. 

Eccentricities were chosen uniformly between 0.01 and 0.05, and inclinations were chosen uniformly between $0 ^\circ$ and $5^\circ$. All other angles were chosen uniformly between 0 and 2$\pi$. Similar to \citet{Anderson19}, we set each planet's radius to $1.6R_J$, assuming the planets are still young, and we set the planet mass to be uniformly sampled between $0.5 M_J$ and $2 M_J$.

We used Rebound to run the N-body scattering simulations \citep{rebound}. We included general relativity correction, using the \texttt{gr-potential} option, and we set the escape distance to 1000 AU. In addition, we assumed a simple prescription in which bodies collide while conserving mass and momentum, as in the built-in REBOUND \texttt{collision} routine. This sticky-sphere prescription ignores alternative collision outcomes to mergers \citep{Jiaru21, Ghosh24}. The radii of the planets are adjusted to conserve volume.


To implement tidal effects efficiently, we assume the planets are quickly circularized when their periastron approaches within 0.03 AU of the host star.  We note that 0.03 AU is chosen so that the eccentricity distribution of the hot Jupiters matches best with observations \citep{Hansen10}. Changing the critical semi-major axis for tides does not qualitatively change our results. If such a close approach occurs, the integration is halted, and we apply angular momentum conservation to determine the planet’s new semi-major axis while setting its eccentricity to zero. After applying the tidal prescription, we continue the integrations for an additional $10^7\,P_{\mathrm{HJ}}$ to allow the systems to relax, where $P_{\mathrm{HJ}}$ is the orbital period of the hot Jupiter. During this phase, tidal interactions with the host star can become significant due to the close proximity to the host star, and we therefore adopt the constant time-lag tidal model implemented in \texttt{REBOUNDx}, using a Love number $k_2 = 0.3$ and a constant time lag of $\tau = 1\,\mathrm{s}$ \citep[e.g.,][]{Anderson16}. The constant time-lag equilibrium tide model provides a good approximation for orbits with moderate to low eccentricities, suitable for the phase after our tidal prescription. We also note that adopting $\tau = 0.1$ or $10,\mathrm{s}$ does not qualitatively affect our results.

If no planets hits our tidal prescription radius, we run each simulation for $10^8 P_{in}$, to incorporate the long-term secular effects, where $P_{in}$ is the period of the innermost planet. We use a time step of $0.001P_{in}$, to resolve the planetary orbits. We adopted the hybrid time-reversible integrator \texttt{TRACE} (\citet{Lu24}) because it employs a hybrid approach, combining WHFast for long-term integrations with BS or IAS15 for close encounters. 






\section{Results}
\label{sec:results}

We ran a total of 1500 three-body scattering simulations on a time scale of $10^8 P_{in}$: 500 each for hot, warm, and cold scattering. If the separation of a planet from its host star reaches within 0.03 AU, the simulation is stopped, and we tidally circularize the planet's orbit (see section \ref{sec:setup} for our tidal prescription). Otherwise, all simulations were run for the full timescale. At the end of this timescale, the orbital parameters were recorded. 




The migration of the planets due to scattering is illustrated in Figure \ref{fig:cartoon}. The figure shows that only a small fraction of the planets from cold scattering ($3.7\%$) and warm scattering ($5.6\%$) form hot Jupiters. Observationally, cold Jupiters are far more abundant than warm and hot Jupiters \citep{Fulton21, Wittenmyer2020},  and thus we expect cold scattering plays a significant role in the formation of hot Jupiters. However, cold scattering rarely ($<1\%$) produces the best-characterized warm Jupiters ($0.1-0.3$AU). These warm Jupiters are primarily produced via hot and warm scattering. 

Additionally, we note that in the hot scattering simulations, 0.4\% of the planets collided with the star, in warm scattering, 0.6\% collided with the star, and in cold scattering, 0.4\% collided with the star. In hot scattering, 8.6\% of the total planets in our 500 simulations were ejected. In warm scattering, 21.7\% were ejected, and in cold scattering, 36.4\%  were ejected. Following ejection, the system ends with two planets most of the time (see section \ref{sec:companions}).

In the following sections, we discuss the specific outcomes for hot Jupiters in section \ref{sec:hotjup} and for warm Jupiters in section \ref{sec:warmjup}, and we discuss the properties of their companions in section \ref{sec:companions}.

\subsection{Hot Jupiter Orbital Distributions}
\label{sec:hotjup}

In our simulations, hot Jupiters, defined as any planet ending with a semi-major axis below 0.1 AU, can form either via in situ hot scattering or via distant scattering followed by tidal circularization and orbital shrinking. In the former, planets are initialized in the hot Jupiter semi-major axis range and undergo collisions or scattering that leave them in the same region. We assume that these hot Jupiters migrated to the hot scattering region during the protoplanetary disk phase before experiencing scattering. In situ hot scattering often leads to merged hot Jupiters that have experienced collisions. These merged hot Jupiters have masses higher than those formed through distant scattering and tidal circularization. This is because mergers preferentially produce higher-mass planets while the lower-mass planet typically moves inward to become a hot Jupiter during scattering (as shown in Figure \ref{fig:massratio}).

\begin{figure*}[htbp]
    \centering   \includegraphics[width=0.8\textwidth]{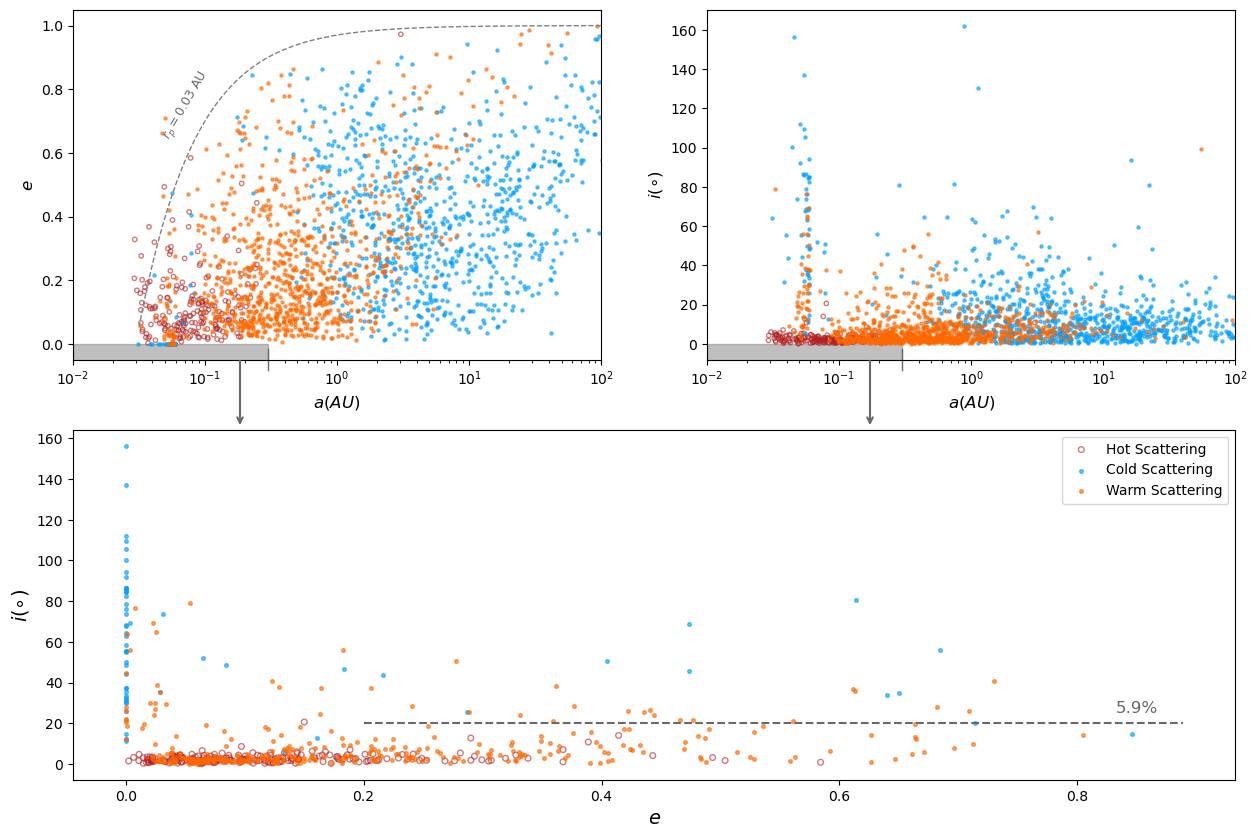}
    \caption{Semi-major axis versus eccentricity and inclination for planets resulting from hot scattering (dark red), warm scattering (orange) and cold scattering (blue). Only 100 of the hot scattering simulations are included for clarity in the hot Jupiter region and to lead to equitable representation of hot Jupiters (since hot scattering produces far more hot Jupiters than warm or cold). In the top left panel, the gray dotted line shows the boundary for a radius of pericenter of 0.03 AU which is used in our tidal prescription.  The bottom panel shows eccentricity versus inclination for all planets with a semi-major axis below 0.3 AU (inner warm and hot Jupiters). 5.9\% of the inner warm Jupiters have eccentricities above 0.2 and inclinations above 20$^\circ$.}
    \label{fig:a_i_e}
\end{figure*}

\begin{figure*}[htbp]
    \centering   \includegraphics[width=0.8\textwidth]{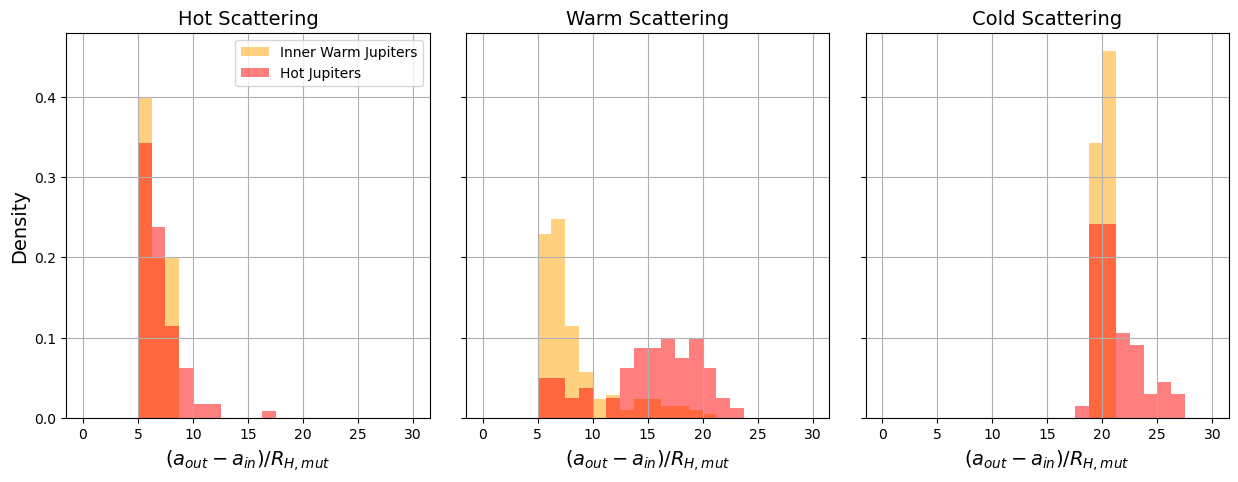}
    \caption{Distributions of the separation in semi-major axis between the inner and outer planet for two-planet systems in terms of their mutual Hill radius from hot, warm, and cold scattering. Only two-planet systems where the inner planet is an inner warm Jupiter ($0.1 < a < 0.3$ AU) or hot Jupiter ($a < 0.1$)
are shown. The red bars indicate hot Jupiters, while the orange bars indicate inner warm Jupiters.}
    \label{fig:rhmut}
\end{figure*}

Figure~\ref{fig:a_i_e} presents the orbital parameters of the planets after scattering, with the percentile breakdown summarized in Table~\ref{tab:perc}. Hot Jupiters produced due to hot scattering are mostly aligned with an inclination below $\sim$20$^{\circ}$. As illustrated in Table \ref{tab:perc}, 90\% of the hot Jupiters produced by hot scatterings have inclination lower than $4.85^\circ$. Hot Jupiters from warm scattering can reach higher inclinations up to $\sim$100$^{\circ}$ (mostly prograde) while cold scattering can produce the most inclined hot Jupiters, even retrograde hot Jupiters, with inclinations as high as 140$^{\circ}$. The eccentricities of hot Jupiters created from cold scattering are nearly  $e \sim 0$ because they have hit our tidal prescription and are instantly assigned a circular orbit. Their eccentricities do not change much during the following phase where the orbits are continued for $10^7P_{HJ}$ because the hot Jupiters resulting from cold scattering typically are isolated with faraway companions (See Figure~\ref{fig:rhmut}). In contrast, hot Jupiters produced from hot scattering can have a broad range of eccentricities ranging from 0 to 0.6 due to the fact that many of them ($\sim 67\%$) do not reach our tidal prescription and instead achieve hot Jupiter status in situ. Hot Jupiters that have experienced our tidal prescription through hot or warm scattering can also acquire a wide range of eccentricities through continued interactions with their companions. Thus, warm scattering can produce hot Jupiters with high obliquities and moderate eccentricities. This contrasts with observations, where currently known misaligned hot Jupiters  are all on circular orbits (X. Wang et al. 2026). This discrepancy suggests that hot Jupiters formed via warm scattering are likely rare comparing to the cold scattering origin. This is consistent with the much larger population of cold Jupiters compared to warm Jupiters.



\begin{table*}[t]
\small
\resizebox{0.8\textwidth}{!}{
\begin{tabular}{|l|c|c|c|c|c|c|}
\hline
 & i$^\circ$, 10\% & i$^\circ$, 20\% & i$^\circ$, 30\% & e, 10\% & e, 20\% & e, 30\%\\ \hline
HS, HJ & 4.85 & 3.95 & 3.16 & 0.28 & 0.19 & 0.15 \\ \hline
HS, WJ & 4.75 & 3.98 & 3.12 & 0.24 & 0.19 & 0.15 \\ \hline
WS, HJ & 53.12 & 37.55 & 27.48 & 0.40 & 0.35 & 0.19 \\ \hline
WS, WJ & 16.08 & 9.33 & 6.95 & 0.47 & 0.34 & 0.27 \\ \hline
WS, CJ & 15.78 & 12.35 & 9.33 & 0.77 & 0.61 & 0.50 \\ \hline
CS, HJ & 99.06 & 85.85 & 77.19 & 0.02 & $< 0.01$ & $< 0.01$ \\ \hline
CS, WJ & 40.71 & 29.21 & 25.47 & 0.73 & 0.66 & 0.62 \\ \hline
CS, CJ & 27.87 & 19.25 & 14.33 & 0.71 & 0.59 & 0.52 \\ \hline
All, HJ & 67.73 & 37.62 & 21.55 & 0.31 & 0.19 & 0.14 \\ \hline
All, WJ & 20.38 & 10.90 & 7.94 & 0.54 & 0.39 & 0.30 \\ \hline
All, CJ & 27.87 & 19.25 & 14.33 & 0.73 & 0.60 & 0.51 \\ \hline
All, inner WJ & 14.30 & 7.21 & 4.90 & 0.48 & 0.33 & 0.24 \\ \hline
All, outer WJ & 23.64 & 13.92 & 9.52 & 0.56 & 0.43 & 0.32 \\ \hline

\end{tabular}
}
\caption{Table with the values corresponding to the top 10\%, 20\% and 30\% of inclination and eccentricity for hot scattering (HS), warm scattering (WS), and cold scattering (CS). Hot Jupiter (HJ), inner warm Jupiter (inner WJ), outer warm Jupiter (outer WJ), and cold Jupiter (CJ) populations are shown. Note that hot Jupiter (HJ) scatterings rarely produce cold Jupiters (CJs), so we do not include a corresponding row for those cases.}
\label{tab:perc}
\end{table*}

\begin{figure}[h]
    \centering   \includegraphics[width=0.48\textwidth]{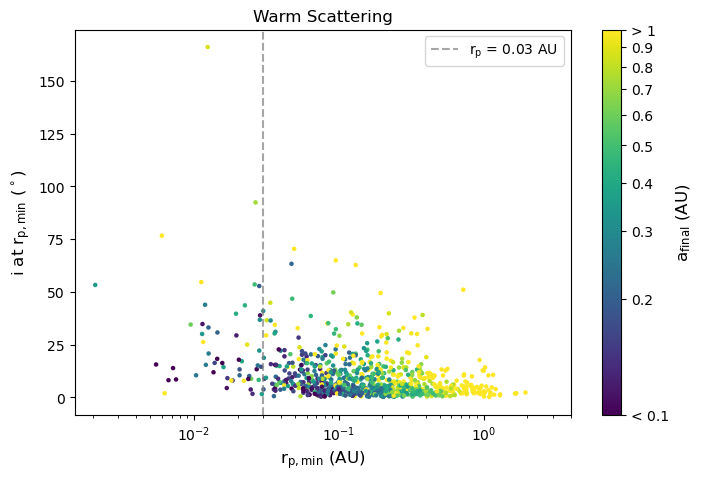}
    \caption{An analysis of the minimum pericenter distance reached during a simulation compared to the inclination at this moment of minimum radius of pericenter, specifically in simulations excluding our tidal prescription. Our tidal distance of 0.03 AU is added for reference. Each point represents a planet from the 500 warm scattering simulations.}
    \label{fig:minrp}
\end{figure}

\subsection{Warm Jupiter Orbital Distributions}
\label{sec:warmjup}
Warm Jupiters, defined as having a semi-major axis between 0.1 and 1 AU, primarily form in situ from warm scattering. Based on Figure \ref{fig:cartoon}, 54.3\% of planets from warm scattering stay in the warm Jupiter range. We divide the warm Jupiter range into two parts: the inner better observationally characterized warm Jupiters (0.1 to 0.3 AU) and the outer warm Jupiters (0.3 to 1 AU). 22.6\% of the planets resulting from warm scattering end up in the inner better characterized region of warm Jupiters. Hot scattering produces warm Jupiters more frequently (30.6\%) than cold scattering (5.3\%). Particularly, cold scattering very rarely produces the best characterized warm Jupiters ($\textless$ 1\%), while hot scattering produces the best characterized warm Jupiters frequently (29.6\%). This occurs because tidal dissipation becomes effective once the outcomes of cold scattering drive planets into the best-characterized warm Jupiter region, which leads to orbital decay into the hot Jupiter region.

Inner warm Jupiters typically have much lower inclinations than outer warm Jupiters as seen in Figure \ref{fig:a_i_e} and in the ``inner WJ'' and ``outer WJ'' rows of Table~\ref{tab:perc}.  Out of all the inner warm Jupiters from warm and cold scattering, only 5.9\% have an inclination above 20$^{\circ}$ and an eccentricity above $e = 0.2$. In contrast, outer warm Jupiters ($\gtrsim$ 0.3 AU) have higher inclinations, with 11.24\% of the planets in the outer warm Jupiter region from warm and cold scattering having an inclination above 20$^\circ$ and eccentricity above e = 0.2. In total, warm Jupiters can have a broad range of eccentricities as high as $e = 0.8$. Therefore, we find that warm Jupiters produced via scattering tend to have a spread in mutual inclinations ($\lesssim 30^\circ$) with moderate to high eccentricities.

The lower inclinations in the inner warm Jupiter region directly correlate to the fact that we include tides in our simulations, unlike many prior planet-planet scattering experiments. In our tidal prescription, as soon as a planet reaches 0.03 AU, we calculate its new semi-major axis via conservation of angular momentum and set its eccentricity to zero. The inclination is kept nearly the same as at the point where the planet reaches 0.03 AU, under perturbations due to their companions. Moreover, higher mutual inclinations often drives further eccentricity excitation due to secular interactions with companions, which can trigger tidal circularization \citep{Bhaskar21}. This process effectively filters out high-inclination warm Jupiters by transforming them into hot Jupiters. Therefore, if we did not apply this tidal prescription, we would expect to see more highly inclined warm Jupiters because they were not ``captured" by our tidal prescription. 

To test this out, we ran warm scattering simulations with no tides and tracked the minimum radius of pericenter that a planet reached and the corresponding inclination at this point of minimum pericenter. In Figure \ref{fig:minrp}, we can see that there are a handful of planets that end up as warm or cold Jupiters with a minimum radius of pericenter below 0.03 AU and a higher inclination. This supports the idea that when including tides, these planets would become misaligned hot Jupiters.

\subsection{Companions}
\label{sec:companions}



Planet-planet scatterings lead most frequently to two-planet systems ($89.2\%$, $72.8\%$ and $64.3\%$ for hot, warm, and cold scattering, respectively). Thus, in this section, we focus on the properties of planetary companions and discuss observational predictions we can make from our scattering models. 

We show in Figure~\ref{fig:rhmut} the separation between the inner and outer planets. We included all the systems that contain a hot/inner warm Jupiter within $0.3$ AU. In both the hot and warm scattering scenarios, the separation ranges around $\sim 5\text{–}10R_{H,\mathrm{mut}}$, and the separation is wider, beyond $\sim 17R_{H,\mathrm{mut}}$ for the systems produced via cold scatterings. For two-planet systems on circular and coplanar orbits, it requires a separation greater than $3.46R_{H,\mathrm{mut}}$ to be stable without close-encounters \citep{Gladman93}, while larger separations are generally needed in the presence of eccentricity and mutual inclination. In three-planet systems, which tend to be more unstable than the two-planet systems, the instability timescale increases exponentially with separation in mutual Hill radii, and reach $\sim$Gyr timescale with a separation of $\sim5.5R_{H,\mathrm{mut}}$ with innermost planet at 3 AU \citep[e.g.,][]{Chatterjee08}. Given the low eccentricities and mutual inclinations in many of the resulting two-planet systems from hot and warm scatterings (see Table \ref{tab:perc}), we expect that many of the resulting systems will be long-term stable, particularly those with larger mutual Hill separations.

The resulting distribution of mutual Hill separations also provides a useful constraint on scattering pathways. Hot scattering produces a large fraction of hot Jupiters with nearby companions ($\sim 5-10\,R_{H,\mathrm{mut}}$), and therefore cannot explain the predominantly isolated nature of the observed hot Jupiter population. This suggests that scattering in the hot Jupiter region is likely rare.

Warm scattering, on the other hand, predominantly produces hot Jupiters with more widely separated companions ($\gtrsim 10\,R_{H,\mathrm{mut}}$) and warm Jupiters with closer companions ($\sim 5\text{--}10\,R_{H,\mathrm{mut}}$). This is consistent with observations, which show that warm Jupiters often host nearby companions \citep{Wu23, Harre24}. Note that only $\sim 5\%$ of warm scattering events lead to hot Jupiter formation, suggesting that hot Jupiters are less commonly produced via this pathway comparing to cold scattering, given the occurrence rate of warm Jupiters is comparable that of hot Jupiters. Finally, while cold scattering rarely produces inner warm Jupiters (see Figure~\ref{fig:cartoon}), it can produce hot Jupiters with much wider companions, consistent with observations of the hot Jupiter population \citep{Bryan16}.

\begin{figure*}[htbp]
    \centering
    \includegraphics[width=1\textwidth]{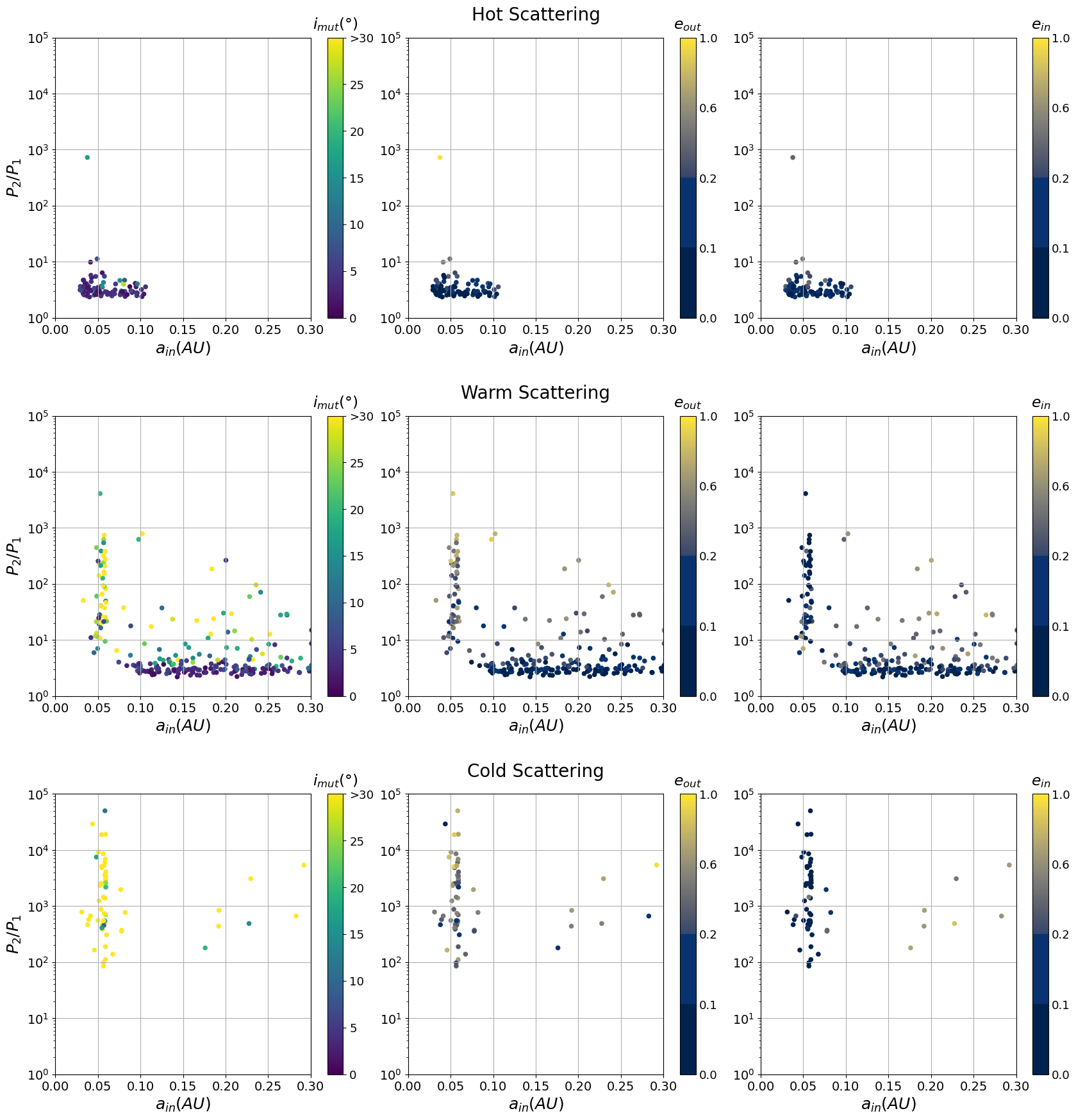}
    \caption{Period ratio versus semi-major axis of the innermost planet with a color bar representing mutual inclination (left-most column), eccentricity of the outer planet (middle column), and eccentricity of the inner planet (right-most column) shown. Only systems with 2 or more planets are accounted for and 1 planet systems are ignored.}
    \label{fig:companion}
\end{figure*}

We tracked the period ratio between planets in the hot, warm, and cold scattering simulations. The period ratio was calculated for all two-planet systems as the outer planet's period over the inner planet's period and for three-planet systems as the middle planet's period over the inner planet's period. One-planet systems are excluded. The results are shown in Figure \ref{fig:companion}.

The top row of Figure \ref{fig:companion} presents results from hot-scattering simulations, while the middle and bottom rows correspond to warm and cold scattering, respectively. We focus solely on inner warm Jupiters and hot Jupiter companions. Most hot-scattering cases produce innermost hot Jupiters with nearby companions (period ratio below 10). In hot scattering, 34\% of systems ended with two hot Jupiters, and 57.4\% ended with a hot and warm Jupiter in the system. In the warm scattering simulations, 5.3\% of hot Jupiters had a warm Jupiter companion and 37.1\% of warm Jupiters had a warm Jupiter companion, implying warm Jupiters typically have nearby companions. Specifically, 147 systems produced multiple warm Jupiters within the same system with warm scattering. No simulations yielded more than one hot Jupiter from the warm scattering simulations. As discussed previously, cold scattering rarely produces warm Jupiters; consequently, none of the cold-scattering simulations formed systems containing both hot and warm Jupiters or multiple warm Jupiters.

It is well-known observationally that hot Jupiters rarely possess nearby companions \citep{Wu23,Harre24}. This is consistent with formation via warm and cold scattering, as illustrated in Figure \ref{fig:companion}. In particular, hot Jupiters produced through cold scattering typically retain companions with period ratios $\gtrsim 200$ and mutual inclinations exceeding $30^\circ$. Those formed through warm scattering generally have companions beyond a period ratio of $\sim 10$, exhibiting moderate mutual inclinations and eccentricities. In contrast, hot scattering tends to produce closer companions with period ratios $\lesssim 100$ and low inclination and eccentricity. This trend differs from most of the observations, which shows that only $\sim10\%$ of the hot Jupiters may contain nearby companions \citep{Wu23}. Thus, this implies that three-planet hot scattering is a less common formation pathway compared to warm or cold scattering.

For completeness, we ran two-body hot scattering simulations to check whether the higher frequency of isolated hot Jupiters can be reproduced. In our two-body hot scattering simulations, we found that ejections were more common than in the three-body case (11.1\% compared to 2.9\%) and that most of the planets did not seem to experience much scattering. The period ratios for most hot Jupiters were still relatively low, meaning two-body scattering does not resolve the issue of fewer isolated hot Jupiters than observed, and we can confirm that hot scattering is not as frequent as warm or cold scattering.




\begin{figure}[h]
    \centering
    \includegraphics[width=0.48\textwidth]{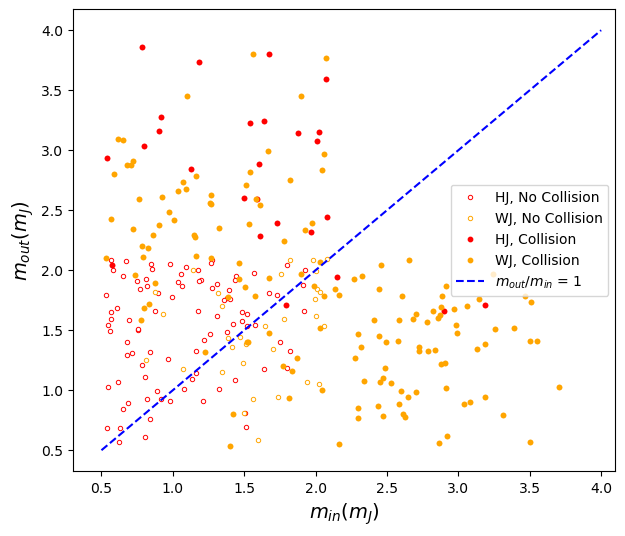}
    \caption{The mass of the inner planet versus the mass of the outer planet for all systems having the innermost planet being a hot Jupiter or inner warm Jupiter resulting from warm and cold scattering simulations. Points with a transparent center indicate a collision. A line for equal inner and outer mass is shown in blue.}
    \label{fig:massratio}
\end{figure}

In contrast, the innermost warm Jupiters are predominantly produced through warm scattering and typically retain nearby companions. As shown in Figure \ref{fig:companion}, these companions exhibit low to moderate eccentricities, generally in the range of 0.1–0.6, and the mutual inclination are typically low $\lesssim 30^\circ$. The right panel further shows that warm Jupiters formed via both warm and cold scattering can attain moderate to high eccentricities.


Figure \ref{fig:massratio} compares the masses between the companion and the innermost planet. The red points correspond to hot Jupiters as the innermost planet in a system, while the orange-yellow points represent systems with an innermost warm Jupiter. The results from warm and cold scattering are combined. Systems that experienced collisions occupy the lower-right and upper-left regions of the diagram, as all planets were initially assigned masses between 0.5 and 2 $M_J$; therefore, any planet exceeding 2 $M_J$ must have undergone a collision. In general, hot Jupiters formed through warm or cold scattering tend to be less massive than their companions, which is consistent with the observation presented by \citep{Zink23}. In contrast, warm Jupiters show no clear mass dependence relative to their companions.

\section{Summary \& Discussion}
\label{sec:summary}

In this paper we revisit planet–planet scattering across three orbital regimes: hot scattering ($a_1=0.03$–$0.1$ AU), warm scattering ($0.1$–$1$ AU), and cold scattering ($1$–$10$ AU), including a simple tidal prescription.  We show that the observed architectures of hot and warm Jupiters are naturally reproduced by a warm-scattering + cold-scattering picture: 

\begin{itemize}
\item \textbf{Warm scattering leaves inner warm Jupiters ($0.1 < a < 0.3$ AU) aligned. }

Warm scattering can produce planets with both high eccentricities and high inclinations. However, systems that reach high inclinations typically also acquire sufficiently large eccentricities so that tides efficiently removes them from the warm Jupiter regime of 0.1-0.3 AU, driving them inward to become hot Jupiters. The surviving warm Jupiters in this best characterized regime are therefore biased toward broad eccentricities but modest inclinations, which match current observations. \citep{Espinoza-Retamal2023}. This is similar to the results of co-planar high eccentricity migration \citep{Li14, Petrovich15}, while warm scattering allows a closer companion compared to the companion required for co-planar high eccentricity migration.

\item \textbf{Warm scattering is required to match current observations.}
  
   Cold Jupiter scattering rarely produces inner warm Jupiters within $\sim 0.3$ AU. So an additional pathway is needed to produce warm Jupiters at $a\simeq0.1$–$0.3$ AU. In our study, warm scattering provides that pathway and produces inner warm Jupiters with the right dynamical characteristics (aligned orbits with broad eccentricity distribution).
  
 \item \textbf{Warm scattering is not the full picture}.
 
\begin{itemize}
 
 \item Warm scattering rarely ($< 1\%$) generates retrograde hot Jupiters. So the retrograde population mostly relies on cold scattering with high-eccentricity migration. 
 
\item Because the observed occurrence rates of warm and hot Jupiters are comparable, warm scattering would require an implausibly larger initial reservoir (by $\sim$1–2 orders of magnitude) to supply the entire hot-Jupiter population, given only $5\%$ of warm scattering become hot Jupiters. A substantial contribution from cold Jupiters is therefore required to reproduce the hot-Jupiter census.

\item Because warm scattering typically results in close-in perturbers capable of exciting the eccentricities of the resulting hot Jupiters, this channel is expected to produce misaligned hot Jupiters with appreciable eccentricities. However, the observed misaligned hot Jupiters are predominantly on circular orbits (X. Wang et al. 2026), suggesting that warm scattering is not the dominant formation pathway for hot Jupiters.

\item Some warm Jupiters host nearby co-transiting companions, including companions near the 2:1 period ratio, which is more naturally explained by disk migration \citep{Laughlin05,Huang16, Trifonov21, Wu23}. This indicates that warm scattering is not guaranteed to operate in every warm-Jupiter system.

\end{itemize}

\item \textbf{Hot scattering should be very rare.} 
Our simulations show that hot scattering tends to leave behind two giant planets on relatively low-eccentricity, low–mutual-inclination orbits with a small period ratio. Although we cannot rule out this pathway, it must be rare in nature: observationally, the close-in companion rate for hot Jupiters is extremely low \citep{Steffen12, Huang16, Wu23}. Even in the few hot-Jupiter systems that do host nearby companions, those companions are almost exclusively small (super-Earth/Neptune-size) planets rather than additional giant planets \citep{Becker15}. WASP-148, which contains a hot Jupiter with a nearby warm-Jupiter companion \citep{Hebrard2020, Wang22}, may be the only convincing example of the type of configuration produced by hot scattering.

\end{itemize}

\textbf{Testable predictions}

If warm scattering plays the significant role proposed here, it should inevitably leave detectable signatures:

\begin{itemize}
    \item \textbf{Warm Jupiters, especially at $a\gtrsim0.3$–$0.4$ AU, should not be exclusively aligned (see Table \ref{tab:perc})}. This can be tested soon with suitable Rossiter–McLaughlin targets, particularly from TESS.
   
    \item \textbf{Warm Jupiters should often have nearby companions with non-negligible mutual inclinations, up to $\lesssim 30^\circ$.} This population is currently underappreciated because RVs are generally insensitive to inclination and transits are intrinsically biased toward coplanar systems. With a combination of TTV/TDV constraints and high-precision RV, these inclined companion architectures should become discoverable and characterizable.
\end{itemize}

\begin{acknowledgments}
The authors thank the anonymous referee for constructive comments that significantly improved the quality of this paper. S.W. gratefully acknowledges insightful discussions with Jiayin Dong, Cristobal Petrovich, Xian-Yu Wang, Malena Rice, and Wenrui Xu. S.W. was supported in part by the NASA Exoplanets Research Program  NNH23ZDA001N-XRP (Grant No. 80NSSC24K0153), the NASA TESS General Investigator Program, Cycle~7, NNNH23ZDA001N-TESS (Grant No. 80NSSC25K7912), and the Heising-Simons Foundation (Grant \#2023-4050). We acknowledge funding support from grant JWST-GO-09025.010-A provided by NASA via the Space Telescope Science Institute under the JWST General Observers Program \#9025. G. Li thanks the support of the Visiting Miller Professorship, funded by the Miller Institute at UC Berkeley.

\end{acknowledgments}

%

\software{\cite{rebound}, \cite{reboundx}, \cite{reboundtrace}}




\bibliography{warmjup}{}

@article{Lu24,
    author = {Lu, Tiger and Hernandez, David M and Rein, Hanno},
    title = {trace: a code for time-reversible astrophysical close encounters},
    journal = {Monthly Notices of the Royal Astronomical Society},
    volume = {533},
    number = {3},
    pages = {3708-3723},
    year = {2024},
    month = {08},
    issn = {0035-8711},
    doi = {10.1093/mnras/stae1982},
    url = {https://doi.org/10.1093/mnras/stae1982},
    eprint = {https://academic.oup.com/mnras/article-pdf/533/3/3708/59026998/stae1982.pdf},
}

@ARTICLE{Xu2024,
       author = {{Xu}, Wenrui and {Wang}, Songhu},
        title = "{Earths Are Not Super-Earths, Saturns Are Not Jupiters: Imprints of Pressure-bump Planet Formation on Planetary Architectures}",
      journal = {\apjl},
         year = 2024,
        month = feb,
       volume = {962},
       number = {1},
          eid = {L4},
        pages = {L4},
          doi = {10.3847/2041-8213/ad1ee1},
archivePrefix = {arXiv},
       eprint = {2401.06217},
 primaryClass = {astro-ph.EP},
       adsurl = {https://ui.adsabs.harvard.edu/abs/2024ApJ...962L...4X}
}

@INCOLLECTION{Triaud2018,
       author = {{Triaud}, Amaury H.~M.~J.},
        title = "{The Rossiter-McLaughlin Effect in Exoplanet Research}",
    booktitle = {Handbook of Exoplanets},
         year = 2018,
       editor = {{Deeg}, Hans J. and {Belmonte}, Juan Antonio},
          eid = {2},
        pages = {2},
          doi = {10.1007/978-3-319-55333-7_2},
       adsurl = {https://ui.adsabs.harvard.edu/abs/2018haex.bookE...2T}
}

@ARTICLE{Fulton2021,
       author = {{Fulton}, Benjamin J. and {Rosenthal}, Lee J. and {Hirsch}, Lea A. and {Isaacson}, Howard and {Howard}, Andrew W. and {Dedrick}, Cayla M. and {Sherstyuk}, Ilya A. and {Blunt}, Sarah C. and {Petigura}, Erik A. and {Knutson}, Heather A. and {Behmard}, Aida and {Chontos}, Ashley and {Crepp}, Justin R. and {Crossfield}, Ian J.~M. and {Dalba}, Paul A. and {Fischer}, Debra A. and {Henry}, Gregory W. and {Kane}, Stephen R. and {Kosiarek}, Molly and {Marcy}, Geoffrey W. and {Rubenzahl}, Ryan A. and {Weiss}, Lauren M. and {Wright}, Jason T.},
        title = "{California Legacy Survey. II. Occurrence of Giant Planets beyond the Ice Line}",
      journal = {\apjs},
         year = 2021,
        month = jul,
       volume = {255},
       number = {1},
          eid = {14},
        pages = {14},
          doi = {10.3847/1538-4365/abfcc1},
archivePrefix = {arXiv},
       eprint = {2105.11584},
 primaryClass = {astro-ph.EP},
       adsurl = {https://ui.adsabs.harvard.edu/abs/2021ApJS..255...14F}
}

@ARTICLE{Hebrard2020,
       author = {{H{\'e}brard}, G. and {D{\'\i}az}, R.~F. and {Correia}, A.~C.~M. and {Collier Cameron}, A. and {Laskar}, J. and {Pollacco}, D. and {Almenara}, J.-M. and {Anderson}, D.~R. and {Barros}, S.~C.~C. and {Boisse}, I. and {Bonomo}, A.~S. and {Bouchy}, F. and {Bou{\'e}}, G. and {Boumis}, P. and {Brown}, D.~J.~A. and {Dalal}, S. and {Deleuil}, M. and {Demangeon}, O.~D.~S. and {Doyle}, A.~P. and {Haswell}, C.~A. and {Hellier}, C. and {Osborn}, H. and {Kiefer}, F. and {Kolb}, U.~C. and {Lam}, K. and {Lecavelier des {\'E}tangs}, A. and {Lopez}, T. and {Martin-Lagarde}, M. and {Maxted}, P. and {McCormac}, J. and {Nielsen}, L.~D. and {Pall{\'e}}, E. and {Prieto-Arranz}, J. and {Queloz}, D. and {Santerne}, A. and {Smalley}, B. and {Turner}, O. and {Udry}, S. and {Verilhac}, D. and {West}, R. and {Wheatley}, P.~J. and {Wilson}, P.~A.},
        title = "{Discovery and characterization of the exoplanets WASP-148b and c. A transiting system with two interacting giant planets}",
      journal = {\aap},
         year = 2020,
        month = aug,
       volume = {640},
          eid = {A32},
        pages = {A32},
          doi = {10.1051/0004-6361/202038296},
archivePrefix = {arXiv},
       eprint = {2004.14645},
 primaryClass = {astro-ph.EP},
       adsurl = {https://ui.adsabs.harvard.edu/abs/2020A&A...640A..32H}
}

@ARTICLE{Espinoza-Retamal2023,
       author = {{Espinoza-Retamal}, Juan I. and {Brahm}, Rafael and {Petrovich}, Cristobal and {Jord{\'a}n}, Andr{\'e}s and {Stef{\'a}nsson}, Gu{\dj}mundur and {Sedaghati}, Elyar and {Hobson}, Melissa J. and {Mu{\~n}oz}, Diego J. and {Boyle}, Gavin and {Leiva}, Rodrigo and {Suc}, Vincent},
        title = "{The Aligned Orbit of the Eccentric Proto Hot Jupiter TOI-3362b}",
      journal = {\apjl},
         year = 2023,
        month = dec,
       volume = {958},
       number = {2},
          eid = {L20},
        pages = {L20},
          doi = {10.3847/2041-8213/ad096d},
archivePrefix = {arXiv},
       eprint = {2309.03306},
 primaryClass = {astro-ph.EP},
       adsurl = {https://ui.adsabs.harvard.edu/abs/2023ApJ...958L..20E}
}

@ARTICLE{Wittenmyer2020,
       author = {{Wittenmyer}, Robert A. and {Wang}, Songhu and {Horner}, Jonathan and {Butler}, R.~P. and {Tinney}, C.~G. and {Carter}, B.~D. and {Wright}, D.~J. and {Jones}, H.~R.~A. and {Bailey}, J. and {O'Toole}, S.~J. and {Johns}, Daniel},
        title = "{Cool Jupiters greatly outnumber their toasty siblings: occurrence rates from the Anglo-Australian Planet Search}",
      journal = {\mnras},
         year = 2020,
        month = feb,
       volume = {492},
       number = {1},
        pages = {377-383},
          doi = {10.1093/mnras/stz3436},
archivePrefix = {arXiv},
       eprint = {1912.01821},
 primaryClass = {astro-ph.EP},
       adsurl = {https://ui.adsabs.harvard.edu/abs/2020MNRAS.492..377W}
}

@article{Anderson19,
    author = {Anderson, Kassandra R and Lai, Dong and Pu, Bonan},
    title = {In situ scattering of warm Jupiters and implications for dynamical histories},
    journal = {Monthly Notices of the Royal Astronomical Society},
    volume = {491},
    number = {1},
    pages = {1369-1383},
    year = {2019},
    month = {11},
    issn = {0035-8711},
    doi = {10.1093/mnras/stz3119},
    url = {https://doi.org/10.1093/mnras/stz3119},
    eprint = {https://academic.oup.com/mnras/article-pdf/491/1/1369/31140979/stz3119.pdf},
}

@article{Rice22_1,
   title={A Tendency Toward Alignment in Single-star Warm-Jupiter Systems},
   volume={164},
   ISSN={1538-3881},
   url={http://dx.doi.org/10.3847/1538-3881/ac8153},
   DOI={10.3847/1538-3881/ac8153},
   number={3},
   journal={The Astronomical Journal},
   publisher={American Astronomical Society},
   author={Rice, Malena and Wang, Songhu and Wang, Xian-Yu and Stefánsson, Guđmundur and Isaacson, Howard and Howard, Andrew W. and Logsdon, Sarah E. and Schweiker, Heidi and Dai, Fei and Brinkman, Casey and Giacalone, Steven and Holcomb, Rae},
   year={2022},
   month=aug, pages={104} }

@misc{Wang24,
      title={Single-Star Warm-Jupiter Systems Tend to Be Aligned, Even Around Hot Stellar Hosts: No $T_{\rm eff}-\lambda$ Dependency}, 
      author={Xian-Yu Wang and Malena Rice and Songhu Wang and Shubham Kanodia and Fei Dai and Sarah E. Logsdon and Heidi Schweiker and Johanna K. Teske and R. Paul Butler and Jeffrey D. Crane and Stephen A. Shectman and Samuel N. Quinn and Veselin B. Kostov and Hugh P. Osborn and Robert F. Goeke and Jason D. Eastman and Avi Shporer and David Rapetti and Karen A. Collins and Cristilyn Watkins and Howard M. Relles and George R. Ricker and Sara Seager and Joshua N. Winn and Jon M. Jenkins},
      year={2024},
      eprint={2408.10038},
      archivePrefix={arXiv},
      primaryClass={astro-ph.EP},
      url={https://arxiv.org/abs/2408.10038}, 
}

@article{Chatterjee08,
   title={Dynamical Outcomes of Planet‐Planet Scattering},
   volume={686},
   ISSN={1538-4357},
   url={http://dx.doi.org/10.1086/590227},
   DOI={10.1086/590227},
   number={1},
   journal={The Astrophysical Journal},
   publisher={American Astronomical Society},
   author={Chatterjee, Sourav and Ford, Eric B. and Matsumura, Soko and Rasio, Frederic A.},
   year={2008},
   month=oct, pages={580–602} }

@article{Juric08,
   title={Dynamical Origin of Extrasolar Planet Eccentricity Distribution},
   volume={686},
   ISSN={1538-4357},
   url={http://dx.doi.org/10.1086/590047},
   DOI={10.1086/590047},
   number={1},
   journal={The Astrophysical Journal},
   publisher={American Astronomical Society},
   author={Jurić, Mario and Tremaine, Scott},
   year={2008},
   month=oct, pages={603–620} }

@article{Petrovich14,
doi = {10.1088/0004-637X/786/2/101},
url = {https://dx.doi.org/10.1088/0004-637X/786/2/101},
year = {2014},
month = {apr},
publisher = {The American Astronomical Society},
volume = {786},
number = {2},
pages = {101},
author = {Petrovich, Cristobal and Tremaine, Scott and Rafikov, Roman},
title = {SCATTERING OUTCOMES OF CLOSE-IN PLANETS: CONSTRAINTS ON PLANET MIGRATION},
journal = {The Astrophysical Journal}
}

@article{Rasio96,
author = {Frederic A. Rasio  and Eric B. Ford },
title = {Dynamical Instabilities and the Formation of Extrasolar Planetary Systems},
journal = {Science},
volume = {274},
number = {5289},
pages = {954-956},
year = {1996},
doi = {10.1126/science.274.5289.954},
URL = {https://www.science.org/doi/abs/10.1126/science.274.5289.954},
eprint = {https://www.science.org/doi/pdf/10.1126/science.274.5289.954}}

@article{Ford08,
   title={Origins of Eccentric Extrasolar Planets: Testing the Planet‐Planet Scattering Model},
   volume={686},
   ISSN={1538-4357},
   url={http://dx.doi.org/10.1086/590926},
   DOI={10.1086/590926},
   number={1},
   journal={The Astrophysical Journal},
   publisher={American Astronomical Society},
   author={Ford, Eric B. and Rasio, Frederic A.},
   year={2008},
   month=oct, pages={621–636} }

@article{Frelikh19,
   title={Signatures of a Planet–Planet Impacts Phase in Exoplanetary Systems Hosting Giant Planets},
   volume={884},
   ISSN={2041-8213},
   url={http://dx.doi.org/10.3847/2041-8213/ab4a7b},
   DOI={10.3847/2041-8213/ab4a7b},
   number={2},
   journal={The Astrophysical Journal Letters},
   publisher={American Astronomical Society},
   author={Frelikh, Renata and Jang, Hyerin and Murray-Clay, Ruth A. and Petrovich, Cristobal},
   year={2019},
   month=oct, pages={L47} }

@article{Fulton21,
   title={California Legacy Survey. II. Occurrence of Giant Planets beyond the Ice Line},
   volume={255},
   ISSN={1538-4365},
   url={http://dx.doi.org/10.3847/1538-4365/abfcc1},
   DOI={10.3847/1538-4365/abfcc1},
   number={1},
   journal={The Astrophysical Journal Supplement Series},
   publisher={American Astronomical Society},
   author={Fulton, Benjamin J. and Rosenthal, Lee J. and Hirsch, Lea A. and Isaacson, Howard and Howard, Andrew W. and Dedrick, Cayla M. and Sherstyuk, Ilya A. and Blunt, Sarah C. and Petigura, Erik A. and Knutson, Heather A. and Behmard, Aida and Chontos, Ashley and Crepp, Justin R. and Crossfield, Ian J. M. and Dalba, Paul A. and Fischer, Debra A. and Henry, Gregory W. and Kane, Stephen R. and Kosiarek, Molly and Marcy, Geoffrey W. and Rubenzahl, Ryan A. and Weiss, Lauren M. and Wright, Jason T.},
   year={2021},
   month=jul, pages={14} }

@ARTICLE{Harre24,
       author = {{Harre}, J. -V. and {Smith}, A.~M.~S. and {Barros}, S.~C.~C. and {Singh}, V. and {Korth}, J. and {Brandeker}, A. and {Collier Cameron}, A. and {Lendl}, M. and {Wilson}, T.~G. and {Borsato}, L. and {Csizmadia}, Sz. and {Cabrera}, J. and {Parviainen}, H. and {Correia}, A.~C.~M. and {Akinsanmi}, B. and {Rosario}, N. and {Leonardi}, P. and {Serrano}, L.~M. and {Alibert}, Y. and {Alonso}, R. and {Asquier}, J. and {B{\'a}rczy}, T. and {Barrado Navascues}, D. and {Baumjohann}, W. and {Benz}, W. and {Billot}, N. and {Broeg}, C. and {Busch}, M. -D. and {Cubillos}, P.~E. and {Davies}, M.~B. and {Deleuil}, M. and {Deline}, A. and {Delrez}, L. and {Demangeon}, O.~D.~S. and {Demory}, B. -O. and {Derekas}, A. and {Edwards}, B. and {Ehrenreich}, D. and {Erikson}, A. and {Fortier}, A. and {Fossati}, L. and {Fridlund}, M. and {Gandolfi}, D. and {Gazeas}, K. and {Gillon}, M. and {G{\"u}del}, M. and {G{\"u}nther}, M.~N. and {Heitzmann}, A. and {Helling}, Ch. and {Isaak}, K.~G. and {Kiss}, L.~L. and {Lam}, K.~W.~F. and {Laskar}, J. and {Lecavelier des Etangs}, A. and {Magrin}, D. and {Maxted}, P.~F.~L. and {Mer{\'\i}n}, B. and {Mordasini}, C. and {Nascimbeni}, V. and {Olofsson}, G. and {Ottensamer}, R. and {Pagano}, I. and {Pall{\'e}}, E. and {Peter}, G. and {Piazza}, D. and {Piotto}, G. and {Pollacco}, D. and {Queloz}, D. and {Ragazzoni}, R. and {Rando}, N. and {Rauer}, H. and {Ribas}, I. and {Santos}, N.~C. and {Scandariato}, G. and {S{\'e}gransan}, D. and {Simon}, A.~E. and {Sousa}, S.~G. and {Stalport}, M. and {Sulis}, S. and {Szab{\'o}}, Gy. M. and {Udry}, S. and {Ulmer}, B. and {Van Grootel}, V. and {Venturini}, J. and {Villaver}, E. and {Viotto}, V. and {Walton}, N.~A. and {West}, R. and {Westerdorff}, K.},
        title = "{Hints of a close outer companion to the ultra-hot Jupiter TOI-2109 b}",
      journal = {\aap},
         year = 2024,
        month = dec,
       volume = {692},
          eid = {A254},
        pages = {A254},
          doi = {10.1051/0004-6361/202451068},
archivePrefix = {arXiv},
       eprint = {2411.07797},
 primaryClass = {astro-ph.EP},
       adsurl = {https://ui.adsabs.harvard.edu/abs/2024A&A...692A.254H}
}

@ARTICLE{Wu23,
       author = {{Wu}, Dong-Hong and {Rice}, Malena and {Wang}, Songhu},
        title = "{Evidence for Hidden Nearby Companions to Hot Jupiters}",
      journal = {\aj},
         year = 2023,
        month = apr,
       volume = {165},
       number = {4},
          eid = {171},
        pages = {171},
          doi = {10.3847/1538-3881/acbf3f},
archivePrefix = {arXiv},
       eprint = {2302.12778},
 primaryClass = {astro-ph.EP},
       adsurl = {https://ui.adsabs.harvard.edu/abs/2023AJ....165..171W}
}

@article{Steffen12,
author = {Jason H. Steffen  and Darin Ragozzine  and Daniel C. Fabrycky  and Joshua A. Carter  and Eric B. Ford  and Matthew J. Holman  and Jason F. Rowe  and William F. Welsh  and William J. Borucki  and Alan P. Boss  and David R. Ciardi  and Samuel N. Quinn },
title = {Kepler constraints on planets near hot Jupiters},
journal = {Proceedings of the National Academy of Sciences},
volume = {109},
number = {21},
pages = {7982-7987},
year = {2012},
doi = {10.1073/pnas.1120970109},
URL = {https://www.pnas.org/doi/abs/10.1073/pnas.1120970109},
eprint = {https://www.pnas.org/doi/pdf/10.1073/pnas.1120970109}}

@ARTICLE{Zink23,
       author = {{Zink}, Jon K. and {Howard}, Andrew W.},
        title = "{Hot Jupiters Have Giant Companions: Evidence for Coplanar High-eccentricity Migration}",
      journal = {\apjl},
         year = 2023,
        month = oct,
       volume = {956},
       number = {1},
          eid = {L29},
        pages = {L29},
          doi = {10.3847/2041-8213/acfdab},
archivePrefix = {arXiv},
       eprint = {2310.01567},
 primaryClass = {astro-ph.EP},
       adsurl = {https://ui.adsabs.harvard.edu/abs/2023ApJ...956L..29Z}
}

@article{Huang16,
doi = {10.3847/0004-637X/825/2/98},
url = {https://doi.org/10.3847/0004-637X/825/2/98},
year = {2016},
month = {jul},
publisher = {The American Astronomical Society},
volume = {825},
number = {2},
pages = {98},
author = {Huang, Chelsea and Wu, Yanqin and Triaud, Amaury H. M. J.},
title = {WARM JUPITERS ARE LESS LONELY THAN HOT JUPITERS: CLOSE NEIGHBORS},
journal = {The Astrophysical Journal}
}

@ARTICLE{Wang21,
       author = {{Wang}, Xian-Yu and {Wang}, Yong-Hao and {Wang}, Songhu and {Wu}, Zhen-Yu and {Rice}, Malena and {Zhou}, Xu and {Hinse}, Tobias C. and {Liu}, Hui-Gen and {Ma}, Bo and {Peng}, Xiyan and {Zhang}, Hui and {Yu}, Cong and {Zhou}, Ji-Lin and {Laughlin}, Gregory},
        title = "{Transiting Exoplanet Monitoring Project (TEMP). VI. The Homogeneous Refinement of System Parameters for 39 Transiting Hot Jupiters with 127 New Light Curves}",
      journal = {\apjs},
         year = 2021,
        month = jul,
       volume = {255},
       number = {1},
          eid = {15},
        pages = {15},
          doi = {10.3847/1538-4365/ac0835},
archivePrefix = {arXiv},
       eprint = {2105.14851},
 primaryClass = {astro-ph.EP},
       adsurl = {https://ui.adsabs.harvard.edu/abs/2021ApJS..255...15W}
}

@article{Batygin16,
doi = {10.3847/0004-637X/829/2/114},
url = {https://doi.org/10.3847/0004-637X/829/2/114},
year = {2016},
month = {sep},
publisher = {The American Astronomical Society},
volume = {829},
number = {2},
pages = {114},
author = {Batygin, Konstantin and Bodenheimer, Peter H. and Laughlin, Gregory P.},
title = {IN SITU FORMATION AND DYNAMICAL EVOLUTION OF HOT JUPITER SYSTEMS},
journal = {The Astrophysical Journal}
}

@ARTICLE{Wu11,
       author = {{Wu}, Yanqin and {Lithwick}, Yoram},
        title = "{Secular Chaos and the Production of Hot Jupiters}",
      journal = {\apj},
         year = 2011,
        month = jul,
       volume = {735},
       number = {2},
          eid = {109},
        pages = {109},
          doi = {10.1088/0004-637X/735/2/109},
archivePrefix = {arXiv},
       eprint = {1012.3475},
 primaryClass = {astro-ph.EP},
       adsurl = {https://ui.adsabs.harvard.edu/abs/2011ApJ...735..109W}
}

@ARTICLE{Naoz11,
       author = {{Naoz}, Smadar and {Farr}, Will M. and {Lithwick}, Yoram and {Rasio}, Frederic A. and {Teyssandier}, Jean},
        title = "{Hot Jupiters from secular planet-planet interactions}",
      journal = {\nat},
         year = 2011,
        month = may,
       volume = {473},
       number = {7346},
        pages = {187-189},
          doi = {10.1038/nature10076},
archivePrefix = {arXiv},
       eprint = {1011.2501},
 primaryClass = {astro-ph.EP},
       adsurl = {https://ui.adsabs.harvard.edu/abs/2011Natur.473..187N}
}

@ARTICLE{Noaz16,
       author = {{Naoz}, Smadar},
        title = "{The Eccentric Kozai-Lidov Effect and Its Applications}",
      journal = {\araa},
         year = 2016,
        month = sep,
       volume = {54},
        pages = {441-489},
          doi = {10.1146/annurev-astro-081915-023315},
archivePrefix = {arXiv},
       eprint = {1601.07175},
 primaryClass = {astro-ph.EP},
       adsurl = {https://ui.adsabs.harvard.edu/abs/2016ARA&A..54..441N}
}

@ARTICLE{Goldreich80,
       author = {{Goldreich}, P. and {Tremaine}, S.},
        title = "{Disk-satellite interactions.}",
      journal = {\apj},
         year = 1980,
        month = oct,
       volume = {241},
        pages = {425-441},
          doi = {10.1086/158356},
       adsurl = {https://ui.adsabs.harvard.edu/abs/1980ApJ...241..425G}
}

@ARTICLE{Lin86,
       author = {{Lin}, D.~N.~C. and {Papaloizou}, John},
        title = "{On the Tidal Interaction between Protoplanets and the Protoplanetary Disk. III. Orbital Migration of Protoplanets}",
      journal = {\apj},
         year = 1986,
        month = oct,
       volume = {309},
        pages = {846},
          doi = {10.1086/164653},
       adsurl = {https://ui.adsabs.harvard.edu/abs/1986ApJ...309..846L}
}

@ARTICLE{Bryan16,
       author = {{Bryan}, Marta L. and {Knutson}, Heather A. and {Howard}, Andrew W. and {Ngo}, Henry and {Batygin}, Konstantin and {Crepp}, Justin R. and {Fulton}, B.~J. and {Hinkley}, Sasha and {Isaacson}, Howard and {Johnson}, John A. and {Marcy}, Geoffry W. and {Wright}, Jason T.},
        title = "{Statistics of Long Period Gas Giant Planets in Known Planetary Systems}",
      journal = {\apj},
         year = 2016,
        month = apr,
       volume = {821},
       number = {2},
          eid = {89},
        pages = {89},
          doi = {10.3847/0004-637X/821/2/89},
archivePrefix = {arXiv},
       eprint = {1601.07595},
 primaryClass = {astro-ph.EP},
       adsurl = {https://ui.adsabs.harvard.edu/abs/2016ApJ...821...89B}
}

@article{Kokubo02,
doi = {10.1086/344105},
url = {https://doi.org/10.1086/344105},
year = {2002},
month = {dec},
publisher = {},
volume = {581},
number = {1},
pages = {666},
author = {Kokubo, Eiichiro and Ida, Shigeru},
title = {Formation of Protoplanet Systems and Diversity of Planetary Systems},
journal = {The Astrophysical Journal}
}

@ARTICLE{Goldreich04,
       author = {{Goldreich}, Peter and {Lithwick}, Yoram and {Sari}, Re'em},
        title = "{Final Stages of Planet Formation}",
      journal = {\apj},
         year = 2004,
        month = oct,
       volume = {614},
       number = {1},
        pages = {497-507},
          doi = {10.1086/423612},
archivePrefix = {arXiv},
       eprint = {astro-ph/0404240},
 primaryClass = {astro-ph},
       adsurl = {https://ui.adsabs.harvard.edu/abs/2004ApJ...614..497G}
}

@ARTICLE{Ida04,
       author = {{Ida}, S. and {Lin}, D.~N.~C.},
        title = "{Toward a Deterministic Model of Planetary Formation. I. A Desert in the Mass and Semimajor Axis Distributions of Extrasolar Planets}",
      journal = {\apj},
         year = 2004,
        month = mar,
       volume = {604},
       number = {1},
        pages = {388-413},
          doi = {10.1086/381724},
archivePrefix = {arXiv},
       eprint = {astro-ph/0312144},
 primaryClass = {astro-ph},
       adsurl = {https://ui.adsabs.harvard.edu/abs/2004ApJ...604..388I}
}

@article{Chambers96,
title = {The Stability of Multi-Planet Systems},
journal = {Icarus},
volume = {119},
number = {2},
pages = {261-268},
year = {1996},
issn = {0019-1035},
doi = {https://doi.org/10.1006/icar.1996.0019},
url = {https://www.sciencedirect.com/science/article/pii/S0019103596900196},
author = {J.E. Chambers and G.W. Wetherill and A.P. Boss}
}

@ARTICLE{Lin97,
       author = {{Lin}, D.~N.~C. and {Ida}, Shigeru},
        title = "{On the Origin of Massive Eccentric Planets}",
      journal = {\apj},
         year = 1997,
        month = mar,
       volume = {477},
       number = {2},
        pages = {781-791},
          doi = {10.1086/303738},
       adsurl = {https://ui.adsabs.harvard.edu/abs/1997ApJ...477..781L}
}

@article{Adams03,
   title={Migration and dynamical relaxation in crowded systems of giant planets},
   volume={163},
   ISSN={0019-1035},
   url={http://dx.doi.org/10.1016/S0019-1035(03)00081-2},
   DOI={10.1016/s0019-1035(03)00081-2},
   number={2},
   journal={Icarus},
   publisher={Elsevier BV},
   author={Adams, Fred C. and Laughlin, Gregory},
   year={2003},
   month=jun, pages={290–306} }

@article{Boss06,
doi = {10.1086/500613},
url = {https://doi.org/10.1086/500613},
year = {2006},
month = {jan},
publisher = {},
volume = {637},
number = {2},
pages = {L137},
author = {Boss, Alan P.},
title = {On the Formation of Gas Giant Planets on Wide Orbits},
journal = {The Astrophysical Journal}
}

@ARTICLE{Scharf09,
       author = {{Scharf}, Caleb and {Menou}, Kristen},
        title = "{Long-Period Exoplanets From Dynamical Relaxation}",
      journal = {\apjl},
         year = 2009,
        month = mar,
       volume = {693},
       number = {2},
        pages = {L113-L117},
          doi = {10.1088/0004-637X/693/2/L113},
archivePrefix = {arXiv},
       eprint = {0811.1981},
 primaryClass = {astro-ph},
       adsurl = {https://ui.adsabs.harvard.edu/abs/2009ApJ...693L.113S}
}

@article{Veras09,
doi = {10.1088/0004-637X/696/2/1600},
url = {https://doi.org/10.1088/0004-637X/696/2/1600},
year = {2009},
month = {apr},
publisher = {The American Astronomical Society},
volume = {696},
number = {2},
pages = {1600},
author = {Veras, Dimitri and Crepp, Justin R. and Ford, Eric B.},
title = {FORMATION, SURVIVAL, AND DETECTABILITY OF PLANETS BEYOND 100 AU},
journal = {The Astrophysical Journal}
}

@article{Mustill17,
    author = {Mustill, Alexander J. and Davies, Melvyn B. and Johansen, Anders},
    title = {The effects of external planets on inner systems: multiplicities, inclinations and pathways to eccentric warm Jupiters},
    journal = {Monthly Notices of the Royal Astronomical Society},
    volume = {468},
    number = {3},
    pages = {3000-3023},
    year = {2017},
    month = {03},
    issn = {0035-8711},
    doi = {10.1093/mnras/stx693},
    url = {https://doi.org/10.1093/mnras/stx693},
    eprint = {https://academic.oup.com/mnras/article-pdf/468/3/3000/13628841/stx693.pdf},
}

@ARTICLE{Gladman93,
       author = {{Gladman}, Brett},
        title = "{Dynamics of Systems of Two Close Planets}",
      journal = {\icarus},
         year = 1993,
        month = nov,
       volume = {106},
       number = {1},
        pages = {247-263},
          doi = {10.1006/icar.1993.1169},
       adsurl = {https://ui.adsabs.harvard.edu/abs/1993Icar..106..247G}
}

@ARTICLE{Trifonov21,
       author = {{Trifonov}, Trifon and {Brahm}, Rafael and {Espinoza}, Nestor and {Henning}, Thomas and {Jord{\'a}n}, Andr{\'e}s and {Nesvorny}, David and {Dawson}, Rebekah I. and {Lissauer}, Jack J. and {Lee}, Man Hoi and {Kossakowski}, Diana and {Rojas}, Felipe I. and {Hobson}, Melissa J. and {Sarkis}, Paula and {Schlecker}, Martin and {Bitsch}, Bertram and {Bakos}, Gaspar {\'A}. and {Barbieri}, Mauro and {Bhatti}, W. and {Butler}, R. Paul and {Crane}, Jeffrey D. and {Nandakumar}, Sangeetha and {D{\'\i}az}, Mat{\'\i}as R. and {Shectman}, Stephen and {Teske}, Johanna and {Torres}, Pascal and {Suc}, Vincent and {Vines}, Jose I. and {Wang}, Sharon X. and {Ricker}, George R. and {Shporer}, Avi and {Vanderburg}, Andrew and {Dragomir}, Diana and {Vanderspek}, Roland and {Burke}, Christopher J. and {Daylan}, Tansu and {Shiao}, Bernie and {Jenkins}, Jon M. and {Wohler}, Bill and {Seager}, Sara and {Winn}, Joshua N.},
        title = "{A Pair of Warm Giant Planets near the 2:1 Mean Motion Resonance around the K-dwarf Star TOI-2202}",
      journal = {\aj},
         year = 2021,
        month = dec,
       volume = {162},
       number = {6},
          eid = {283},
        pages = {283},
          doi = {10.3847/1538-3881/ac1bbe},
archivePrefix = {arXiv},
       eprint = {2108.05323},
 primaryClass = {astro-ph.EP},
       adsurl = {https://ui.adsabs.harvard.edu/abs/2021AJ....162..283T}
}

@ARTICLE{Wang22,
       author = {{Wang}, Xian-Yu and {Rice}, Malena and {Wang}, Songhu and {Pu}, Bonan and {Stef{\'a}nsson}, Gudmundur and {Mahadevan}, Suvrath and {Radzom}, Brandon and {Giacalone}, Steven and {Wu}, Zhen-Yu and {Esposito}, Thomas M. and {Dalba}, Paul A. and {Avsar}, Arin and {Holden}, Bradford and {Skiff}, Brian and {Polakis}, Tom and {Voeller}, Kevin and {Logsdon}, Sarah E. and {Klusmeyer}, Jessica and {Schweiker}, Heidi and {Wu}, Dong-Hong and {Beard}, Corey and {Dai}, Fei and {Lubin}, Jack and {Weiss}, Lauren M. and {Bender}, Chad F. and {Blake}, Cullen H. and {Dressing}, Courtney D. and {Halverson}, Samuel and {Hearty}, Fred and {Howard}, Andrew W. and {Huber}, Daniel and {Isaacson}, Howard and {Jackman}, James A.~G. and {Llama}, Joe and {McElwain}, Michael W. and {Rajagopal}, Jayadev and {Roy}, Arpita and {Robertson}, Paul and {Schwab}, Christian and {Shkolnik}, Evgenya L. and {Wright}, Jason T. and {Laughlin}, Gregory},
        title = "{The Aligned Orbit of WASP-148b, the Only Known Hot Jupiter with a nearby Warm Jupiter Companion, from NEID and HIRES}",
      journal = {\apjl},
         year = 2022,
        month = feb,
       volume = {926},
       number = {2},
          eid = {L8},
        pages = {L8},
          doi = {10.3847/2041-8213/ac4f44},
archivePrefix = {arXiv},
       eprint = {2110.08832},
 primaryClass = {astro-ph.EP},
       adsurl = {https://ui.adsabs.harvard.edu/abs/2022ApJ...926L...8W}
}

@ARTICLE{Laughlin05,
       author = {{Laughlin}, Gregory and {Butler}, R. Paul and {Fischer}, Debra A. and {Marcy}, Geoffrey W. and {Vogt}, Steven S. and {Wolf}, Aaron S.},
        title = "{The GJ 876 Planetary System: A Progress Report}",
      journal = {\apj},
         year = 2005,
        month = apr,
       volume = {622},
       number = {2},
        pages = {1182-1190},
          doi = {10.1086/424686},
archivePrefix = {arXiv},
       eprint = {astro-ph/0407441},
 primaryClass = {astro-ph},
       adsurl = {https://ui.adsabs.harvard.edu/abs/2005ApJ...622.1182L}
}

@ARTICLE{Bhaskar21,
       author = {{Bhaskar}, Hareesh and {Li}, Gongjie and {Hadden}, Sam and {Payne}, Matthew J. and {Holman}, Matthew J.},
        title = "{Mildly Hierarchical Triple Dynamics and Applications to the Outer Solar System}",
      journal = {\aj},
         year = 2021,
        month = jan,
       volume = {161},
       number = {1},
          eid = {48},
        pages = {48},
          doi = {10.3847/1538-3881/abcbfc},
archivePrefix = {arXiv},
       eprint = {2008.04335},
 primaryClass = {astro-ph.EP},
       adsurl = {https://ui.adsabs.harvard.edu/abs/2021AJ....161...48B}
}

@ARTICLE{Anderson16,
       author = {{Anderson}, Kassandra R. and {Storch}, Natalia I. and {Lai}, Dong},
        title = "{Formation and stellar spin-orbit misalignment of hot Jupiters from Lidov-Kozai oscillations in stellar binaries}",
      journal = {\mnras},
         year = 2016,
        month = mar,
       volume = {456},
       number = {4},
        pages = {3671-3701},
          doi = {10.1093/mnras/stv2906},
archivePrefix = {arXiv},
       eprint = {1510.08918},
 primaryClass = {astro-ph.EP},
       adsurl = {https://ui.adsabs.harvard.edu/abs/2016MNRAS.456.3671A}
}

@ARTICLE{Albrecht22,
       author = {{Albrecht}, Simon H. and {Dawson}, Rebekah I. and {Winn}, Joshua N.},
        title = "{Stellar Obliquities in Exoplanetary Systems}",
      journal = {\pasp},
         year = 2022,
        month = aug,
       volume = {134},
       number = {1038},
          eid = {082001},
        pages = {082001},
          doi = {10.1088/1538-3873/ac6c09},
archivePrefix = {arXiv},
       eprint = {2203.05460},
 primaryClass = {astro-ph.EP},
       adsurl = {https://ui.adsabs.harvard.edu/abs/2022PASP..134h2001A}
}

@ARTICLE{Hansen10,
       author = {{Hansen}, Brad M.~S.},
        title = "{Calibration of Equilibrium Tide Theory for Extrasolar Planet Systems}",
      journal = {\apj},
         year = 2010,
        month = nov,
       volume = {723},
       number = {1},
        pages = {285-299},
          doi = {10.1088/0004-637X/723/1/285},
archivePrefix = {arXiv},
       eprint = {1009.3027},
 primaryClass = {astro-ph.SR},
       adsurl = {https://ui.adsabs.harvard.edu/abs/2010ApJ...723..285H}
}

@ARTICLE{Petrovich15,
       author = {{Petrovich}, Cristobal},
        title = "{Hot Jupiters from Coplanar High-eccentricity Migration}",
      journal = {\apj},
         year = 2015,
        month = may,
       volume = {805},
       number = {1},
          eid = {75},
        pages = {75},
          doi = {10.1088/0004-637X/805/1/75},
archivePrefix = {arXiv},
       eprint = {1409.8296},
 primaryClass = {astro-ph.EP},
       adsurl = {https://ui.adsabs.harvard.edu/abs/2015ApJ...805...75P}
}

@ARTICLE{Li14,
       author = {{Li}, Gongjie and {Naoz}, Smadar and {Kocsis}, Bence and {Loeb}, Abraham},
        title = "{Eccentricity Growth and Orbit Flip in Near-coplanar Hierarchical Three-body Systems}",
      journal = {\apj},
         year = 2014,
        month = apr,
       volume = {785},
       number = {2},
          eid = {116},
        pages = {116},
          doi = {10.1088/0004-637X/785/2/116},
archivePrefix = {arXiv},
       eprint = {1310.6044},
 primaryClass = {astro-ph.EP},
       adsurl = {https://ui.adsabs.harvard.edu/abs/2014ApJ...785..116L}
}

@ARTICLE{Becker15,
       author = {{Becker}, Juliette C. and {Vanderburg}, Andrew and {Adams}, Fred C. and {Rappaport}, Saul A. and {Schwengeler}, Hans Martin},
        title = "{WASP-47: A Hot Jupiter System with Two Additional Planets Discovered by K2}",
      journal = {\apjl},
         year = 2015,
        month = oct,
       volume = {812},
       number = {2},
          eid = {L18},
        pages = {L18},
          doi = {10.1088/2041-8205/812/2/L18},
archivePrefix = {arXiv},
       eprint = {1508.02411},
 primaryClass = {astro-ph.EP},
       adsurl = {https://ui.adsabs.harvard.edu/abs/2015ApJ...812L..18B}
}

@ARTICLE{Canas19,
       author = {{Ca{\~n}as}, Caleb I. and {Wang}, Songhu and {Mahadevan}, Suvrath and {Bender}, Chad F. and {De Lee}, Nathan and {Fleming}, Scott W. and {Garc{\'\i}a-Hern{\'a}ndez}, D.~A. and {Hearty}, Fred R. and {Majewski}, Steven R. and {Roman-Lopes}, Alexandre and {Schneider}, Donald P. and {Stassun}, Keivan G.},
        title = "{Kepler-730: A Hot Jupiter System with a Close-in, Transiting, Earth-sized Planet}",
      journal = {\apjl},
         year = 2019,
        month = jan,
       volume = {870},
       number = {2},
          eid = {L17},
        pages = {L17},
          doi = {10.3847/2041-8213/aafa1e},
archivePrefix = {arXiv},
       eprint = {1812.08358},
 primaryClass = {astro-ph.EP},
       adsurl = {https://ui.adsabs.harvard.edu/abs/2019ApJ...870L..17C}
}

@ARTICLE{Goldreich03,
       author = {{Goldreich}, Peter and {Sari}, Re'em},
        title = "{Eccentricity Evolution for Planets in Gaseous Disks}",
      journal = {\apj},
         year = 2003,
        month = mar,
       volume = {585},
       number = {2},
        pages = {1024-1037},
          doi = {10.1086/346202},
archivePrefix = {arXiv},
       eprint = {astro-ph/0202462},
 primaryClass = {astro-ph},
       adsurl = {https://ui.adsabs.harvard.edu/abs/2003ApJ...585.1024G}
}

@ARTICLE{Winn15,
       author = {{Winn}, Joshua N. and {Fabrycky}, Daniel C.},
        title = "{The Occurrence and Architecture of Exoplanetary Systems}",
      journal = {\araa},
         year = 2015,
        month = aug,
       volume = {53},
        pages = {409-447},
          doi = {10.1146/annurev-astro-082214-122246},
archivePrefix = {arXiv},
       eprint = {1410.4199},
 primaryClass = {astro-ph.EP},
       adsurl = {https://ui.adsabs.harvard.edu/abs/2015ARA&A..53..409W}
}

@article{Matsumura10,
doi = {10.1088/0004-637X/714/1/194},
url = {https://doi.org/10.1088/0004-637X/714/1/194},
year = {2010},
month = {apr},
publisher = {The American Astronomical Society},
volume = {714},
number = {1},
pages = {194},
author = {Matsumura, Soko and Thommes, Edward W. and Chatterjee, Sourav and Rasio, Frederic A.},
title = {UNSTABLE PLANETARY SYSTEMS EMERGING OUT OF GAS DISKS},
journal = {The Astrophysical Journal}
}

@article{Jiaru21,
    author = {Li, Jiaru and Lai, Dong and Anderson, Kassandra R and Pu, Bonan},
    title = {Giant planet scatterings and collisions: hydrodynamics, merger-ejection branching ratio, and properties of the remnants},
    journal = {Monthly Notices of the Royal Astronomical Society},
    volume = {501},
    number = {2},
    pages = {1621-1632},
    year = {2020},
    month = {12},
    issn = {0035-8711},
    doi = {10.1093/mnras/staa3779},
    url = {https://doi.org/10.1093/mnras/staa3779},
    eprint = {https://academic.oup.com/mnras/article-pdf/501/2/1621/35327120/staa3779.pdf},
}

@article{Ghosh24,
doi = {10.3847/1538-3881/ad7d8a},
url = {https://doi.org/10.3847/1538-3881/ad7d8a},
year = {2024},
month = {nov},
publisher = {The American Astronomical Society},
volume = {168},
number = {6},
pages = {238},
author = {Ghosh, Tuhin and Chatterjee, Sourav and Lombardi, James C.},
title = {Outcomes of Sub-Neptune Collisions},
journal = {The Astronomical Journal}
}

@ARTICLE{rebound,
       author = {{Rein}, H. and {Liu}, S. -F.},
        title = "{REBOUND: an open-source multi-purpose N-body code for collisional dynamics}",
      journal = {\aap},
         year = 2012,
        month = jan,
       volume = {537},
          eid = {A128},
        pages = {A128},
          doi = {10.1051/0004-6361/201118085},
archivePrefix = {arXiv},
       eprint = {1110.4876},
 primaryClass = {astro-ph.EP},
       adsurl = {https://ui.adsabs.harvard.edu/abs/2012A&A...537A.128R}
}

@ARTICLE{reboundx,
       author = {{Tamayo}, Daniel and {Rein}, Hanno and {Shi}, Pengshuai and {Hernandez}, David M.},
        title = "{REBOUNDx: a library for adding conservative and dissipative forces to otherwise symplectic N-body integrations}",
      journal = {\mnras},
         year = 2020,
        month = jan,
       volume = {491},
       number = {2},
        pages = {2885-2901},
          doi = {10.1093/mnras/stz2870},
archivePrefix = {arXiv},
       eprint = {1908.05634},
 primaryClass = {astro-ph.EP},
       adsurl = {https://ui.adsabs.harvard.edu/abs/2020MNRAS.491.2885T}
}

@ARTICLE{reboundtrace,
        author = {{Lu}, Tiger and {Hernandez}, David M. and {Rein}, Hanno},
        title = "{TRACE: a code for time-reversible astrophysical close encounters}",
      journal = {\mnras},
         year = 2024,
        month = sep,
       volume = {533},
       number = {3},
        pages = {3708-3723},
          doi = {10.1093/mnras/stae1982},
archivePrefix = {arXiv},
       eprint = {2405.03800},
 primaryClass = {astro-ph.EP},
       adsurl = {https://ui.adsabs.harvard.edu/abs/2024MNRAS.533.3708L}
}
\bibliographystyle{aasjournalv7}

\end{document}